COSPAR02-A-00895

# Developing Basic Space Science World Wide: Progress Report

Hans J. Haubold [1] and Willem Wamsteker [2]

[1] *Office for Outer Space Affairs, United Nations, Vienna International Centre, P.O. Box 500, 1400 Vienna, Austria*
[2] *European Space Agency, Satellite Tracking Station Vilspa, P.O. Box 50727, 28080 Madrid, Spain*

## Abstract

The UN/ESA Workshops on Basic Space Science is a long-term effort for the development of astronomy and regional and international co-operation in this field on a world wide basis, particularly in developing nations. The first four workshops in this series (India 1991, Costa Rica and Colombia 1992, Nigeria 1993, and Egypt 1994) addressed the status of astronomy in Asia and the Pacific, Latin America and the Caribbean, Africa, and Western Asia, respectively. One major recommendation that emanated from the first four workshops was that small astronomical facilities should be established in developing nations for research and education programmes at the university level and that such facilities should be networked. Subsequently, material for teaching and observing programmes for small optical telescopes were developed or recommended and astronomical telescope facilities have been inaugurated at UN/ESA Workshops on Basic Space Science in Sri Lanka (1995), Honduras (1997), and Jordan (1999). UN/ESA Workshops on Basic Space Science in Germany (1996), France (2000), Mauritius (2001), and Argentina (2002) emphasised the particular importance of astrophysical data systems and the virtual observatory concept for the development of astronomy on a world wide basis. Since 1996, the workshops are contributing to the development of the World Space Observatory concept. Achievements of the series of workshops are briefly summarised in this report.

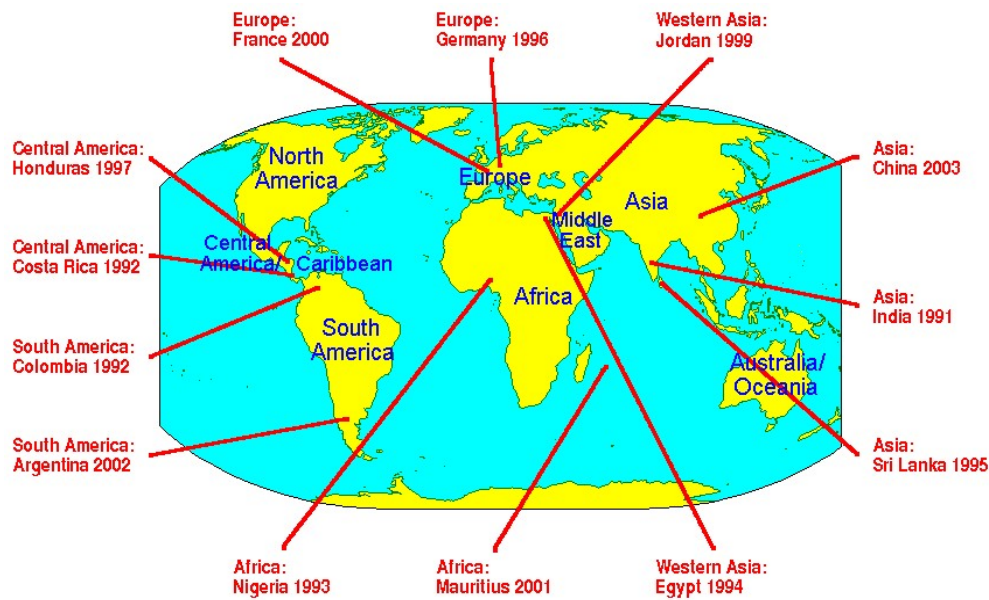

**Access to astronomical literature through the NASA Astrophysical Data System**

Space scientists need, on a day-to-day basis, access to scientific literature, published in journals, proceedings, and books world wide. The ADS is the search system of choice, particularly for astronomers world wide (Wamsteker et al., 2003). The searchable database contains over 2.5 million records. In addition the ADS has over 2 million scanned article pages from about 270,000 articles, dating back as far as 1829. There are currently more than 10,000 regular users. ADS users issue almost 1 million queries per month and receive 30 million records and 1.2 million scanned article pages per month. The ADS is accessed from almost 100 countries with a wide range in number of queries per country. Approximately 1/3 of the use is from the USA, 1/3 from Europe, and 1/3 from other regions of the world. In order to improve access from different parts of the world, ADS maintains 9 mirror sites in Brazil, Chile, China, England, France, Germany, India, Japan, and Russia. Automatic procedures facilitate keeping these mirror sites up-to-date over the network. Both the search system and the scanned articles in ADS can be accessed through email. Email can be used by users that are on slow r unreliable Internet connections. It allows access to ADS for people who do not have a connection that is good enough to use a world-wide-web browser. ADS is currently in the process of developing a stand-alone ADS system that can be updated through Digital Video Discs (DVDs). This would provide access to the capabilities of ADS from sites that do not have any Internet access at all. The capacity of hard disk drives is sufficiently large by now to store a complete ADS system on one large disk.

**Archives, databases, and the emerging virtual observatories**

Latest since the UN/ESA Workshop in Germany (1996), information abundance from and ubiquitous networking of ground-based and space-borne astronomical facilities became immanent issues of the workshop programmes. More and more web-based, distributed research environments for astronomy with massive data sets became available.

An information overload with more efficient ground-based and space-borne astronomical instruments, larger detectors, multiple wavelength coverage and an increasing number of ground and space-based facilities is currently reached. At the same time, information technology is catching up and allows to be able to use the data regardless of where they reside. An increasing need of multi-wavelength observations to understand the underlying physics of the observed phenomena, coupled with availability of formal and informal data archives and the realisation that peta byte databases present a particular challenge to mine valuable data from them, is leading to a federation of archives to set standards, a co-operation among computer scientists and astronomers to create the infrastructure that will lead to a virtual observatory: A cyberspace location where the data is ready to be analysed (Wamsteker et al., 2003).

Given the high cost of modern astronomical observing facilities it is evident that efforts must be made to optimally exploit the data in order to maximise the return on investment. This concept was first implemented on a large scale for the International Ultraviolet Explorer mission and subsequently by Hubble Space Telescope, and has since been taken over for other space-borne and large ground-based facilities. The European HST Science Data Archive is located at the European Southern Observatory (ESO) (Wamsteker et al., 2003). It has been extended to include data from ESO telescopes and instruments, especially the Very Large Telescope (VLT) and Wide Field Imager (WFI). It was thus natural to design the archive such that queries could be extended across its full content, regardless of the origin of the data. This constituted a first step towards a virtual observatory. The ASTROVIRTEL program, first established in 1999-2000 with funding provided by the European Commission, makes it possible for scientists to use this facility for their investigations. At the same time it allowed to establish science requirements for archive cross queries, and to define capabilities required for virtual observatories.

**Conclusion**

Over the period of time from 1991 to 2002, close to 1200 scientists from 124 countries have participated at or contributed to the series of workshops. The UN/ESA Workshops in Mauritius (2001) and in Argentina (2002) have undertaken a detailed assessment of the achievements of the workshop series (Wamsteker et al., 2003). This assessment will be concluded and published at the forthcoming UN/ESA Workshop on Basic Space Science, Beijing, P.R. China, 8-12 September 2003.

**References**

Hill, H.A. and Kroll, R.J., Long-term solar variability and solar seismology, AIP Conference Proceedings, American Institute of Physics, New York, 1992, pp. 170-180.
Wamsteker, W., Albrecht, R., and Haubold, H.J., Developing Basic Space Science World Wide: A Decade of UN/ESA Workshops, Kluwer Academic Publishers, Dordrecht, 2003.